\documentclass[conference]{IEEEtran}
\IEEEoverridecommandlockouts
\usepackage{cite}
\usepackage{amsmath,amssymb,amsfonts}
\usepackage{algorithmic}
\usepackage{graphicx}
\usepackage{textcomp}
\usepackage{xcolor}
\usepackage{xspace}
\usepackage{hyperref}

\newcommand{\rac}{\texttt{RAC}\xspace}

\def\BibTeX{{\rm B\kern-.05em{\sc i\kern-.025em b}\kern-.08em
    T\kern-.1667em\lower.7ex\hbox{E}\kern-.125emX}}

\begin{document}

\title{Random Adaptive Cache Placement Policy\\
}

\author{
Vrushank Ahire\textsuperscript{1*}, Pranav Menon\textsuperscript{1*}, Aniruddh Muley\textsuperscript{2*}, Abhinandan S. Prasad\textsuperscript{1}\\
\textsuperscript{1}Department of Computer Science and Engineering, IIT Ropar, Punjab, India\\
\textsuperscript{2}Department of Mathematics, IIT Ropar, Punjab, India\\
\{2022csb1002, 2022csb1329, 2022mcb1257, abhinandansp\}@iitrpr.ac.in
}

\maketitle

\begin{abstract}
This paper presents a new hybrid cache replacement algorithm that combines random allocation with a modified V-Way cache implementation. Our \rac adapts to complex cache access patterns and optimizes cache usage by improving the utilization of cache sets, unlike traditional cache policies. The algorithm utilizes a $16$-way set-associative cache with $2048$ sets, incorporating dynamic allocation and flexible tag management. \rac extends the V-Way cache design and its variants by optimizing tag and data storage for enhanced efficiency.

We evaluated the algorithm using the ChampSim simulator with four diverse benchmark traces and observed significant improvements in cache hit rates up to $80.82\%$ hit rate. Although the improvements in the instructions per cycle (IPC) were moderate, our findings emphasize the algorithm's potential to enhance cache utilization and reduce memory access times.

%We discuss the algorithm's design, implementation, evaluation results, and potential areas for future optimization, setting the stage for more efficient cache techniques across different computational domains. (444.namd, 445.gobmk, 473.astar, 605.mcf\_s) , with 473.astar showing an
\end{abstract}

\section{Introduction}
%Cache management plays a crucial role in optimizing computer system performance. 
Traditional cache replacement policies often prove inadequate in capturing complex access patterns, resulting in sub-optimal cache utilization wherein infrequently accessed elements persist while more frequently used elements are evicted. 

The V-Way cache \cite{qureshi2005v} was previously proposed to address the challenge of inadequate utilization of some cache sets over others by dynamically adjusting associativity to handle non-uniform access patterns better. The V-Way cache allows for variable associativity per set by decoupling the tag and data stores. It employs a global replacement policy to prioritize data lines based on reuse frequency. However, the performance depends on the optimal tag-to-data ratio (TDR), and TDR varies based on application demands. Further, TDR requires careful tuning to achieve peak performance. %This suggests that there is room for further optimization to realize the potential performance gains fully.

This paper introduces a random adaptive cache (\rac) replacement algorithm designed to address the limitations of the V-Way cache. Our approach incorporates a random eviction policy that simplifies cache management and reduces the complexity of fine-tuning TDR settings. Our method enhances flexibility and improves overall cache efficiency by employing a hybrid design with random allocation in a $16$-way set-associative cache.

% Add the following:
% \begin{itemize}
% \item What problem V-Cache will solve?
% \item Add two or three sentences on V-Cache
% \item Limitations of V-Cache
% \end{itemize}

% While our proposed algorithm demonstrates promising results in improving cache hit rates, it is important to note its limitations. The current implementation introduces additional complexity in cache management, which may result in increased overhead. 

%The subsequent sections of this paper will provide a detailed examination of our cache replacement algorithm's design, present our evaluation methodology and results, and discuss future research directions in this domain.

\section{Design}

% Our cache replacement algorithm is a novel and efficient approach to cache management. It addresses specific challenges in modern cache systems, and their synthesis results in flexible and adaptive solutions.

\begin{figure}[ht]
    \centering
    \includegraphics[width=0.48\textwidth]{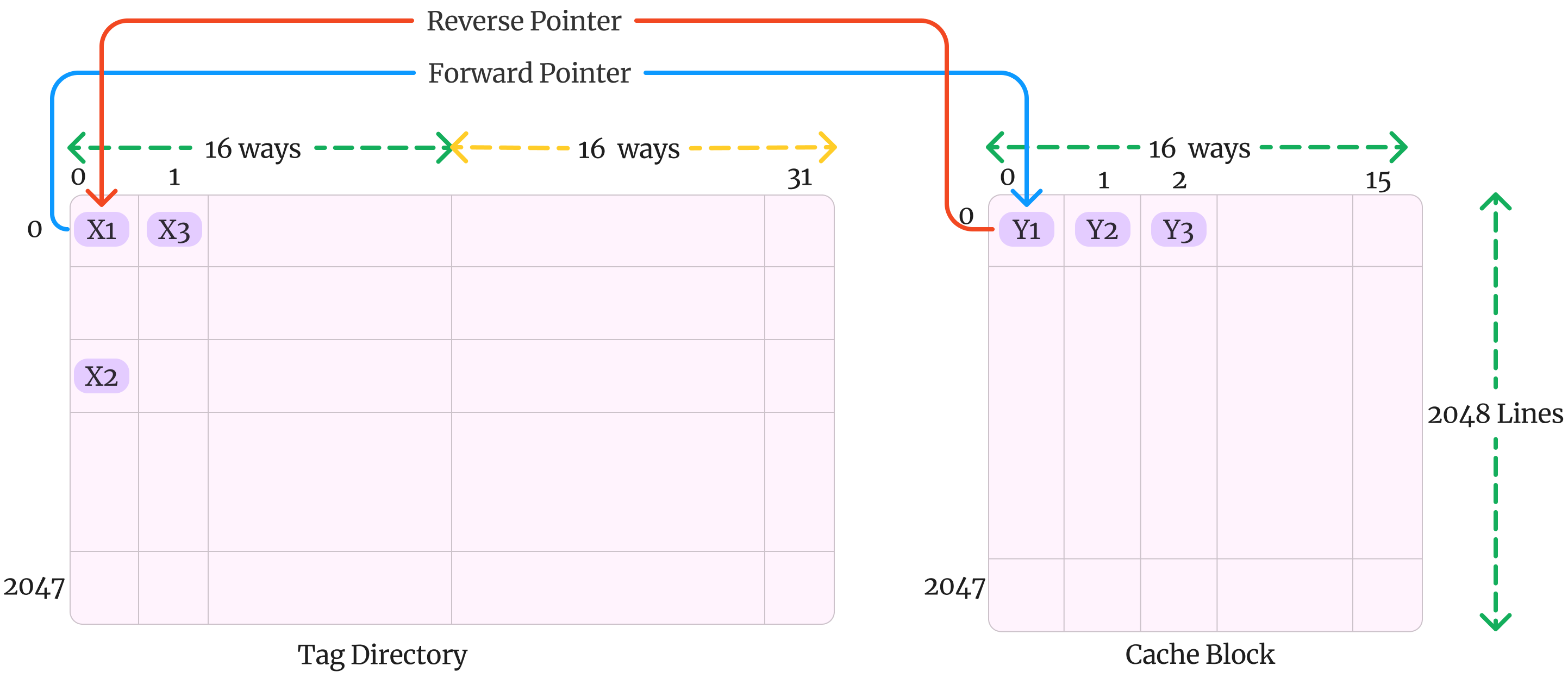}
    \caption{Visualization of Random Adaptive Cache Implementation}
    \label{fig:v_way_cache}
\end{figure}

The proposed design features (as shown in Fig.~\ref{fig:v_way_cache}) a tag directory with $32$ ways is present alongside a cache block with $16$ ways that stores data elements. Each entry in the tag directory contains a forward pointer (blue arrow), while each entry in the cache has a reverse pointer (red arrow). Tags are stored within the tag directory, while the corresponding data elements are stored in the cache.

\subsection{\rac eviction scheme}
\rac utilizes a hybrid scheme that combines a random eviction policy with a few changes to the V-way cache design. Figure \ref{fig:v_way_cache} illustrates that the scheme employs a random eviction policy with the help of a $32-$way tag directory and a $16-$way data directory. When the corresponding tag directory set has space available, but the data directory is full, a random element from the data directory is evicted to make space for a new element. The new element will be added in two steps.

\begin{itemize}
    \item If the data table is full and there is space in the tag directory table for the corresponding set, an element is randomly chosen for eviction from the cache data block. The corresponding tag to this element can be found using the reverse pointer (see Fig. \ref{fig:v_way_cache}) stored in the cache block. This corresponding valid bit for the tag entry is made invalid, creating space for the new element to be added to the cache block.
    \item The tag of the new element is stored in its corresponding set in the tag directory, and the data is entered into the vacated cache block using the forward pointer (see Fig. \href{1}{\ref{fig:v_way_cache}}).
\end{itemize}
% \begin{itemize}
%     \item There are 16 ways of data cache directly linked to a tag directory table that manages data allocation to the cache blocks. \rac allocates memory tags to the new elements. 

%     \item Tag directory table with 16 extra ways to handle frequently used sets: The extra elements can be stored in the allocated 16 ways.
%     % \item Linked list structure for efficiently tracking overflow entries: This enables quick access and management of dynamically allocated cache lines.

% \end{itemize}

This approach allows the cache to maintain a stable base configuration while adapting to sudden spikes in data access or changes in access patterns. The shared pool mechanism ensures that sets experiencing high demand can temporarily expand their capacity, reducing the likelihood of cache thrashing and improving overall hit rates.

\subsection{\rac implementation}
\rac implementation optimizes tag and data storage, further enhancing the efficiency of our cache replacement algorithm. This component consists of two main structures: a tag table with 2048 lines and 32 ways, which allows for tackling a larger number of cache lines per set, and a data table, which has 2048 lines and 16 ways, which store the actual cached data.

This tag and data storage separation allows for more flexible cache management and potentially higher associativity. Our implementation handles four distinct cases for cache fills and evictions:

\begin{enumerate}
    \item When both the tag table index set and the data table have space, the entry is added to both tables.
    \item If the tag table index set is full (32 ways), the least recently used (LRU) policy is employed to evict an element from the tag table. Appropriate changes are made to the data table as well.
    \item When the tag table index set has space but the data table is full, a data element is randomly evicted from the data table to accommodate the new element.
    \item In cases where both the tag table index set and the data table are full, LRU is applied to the tag table index set, and a random eviction is performed on the data table to store the new data and tag.
\end{enumerate}

This approach allows for more efficient use of cache storage and provides flexibility in managing cache contents based on access patterns and storage availability. However, extra hardware is needed to manage the tag directory table.

\section{Evaluation}
To assess the effectiveness of \rac, we conducted evaluations using the ChampSim simulator~\cite{gober2022championship}. Our test bed comprised four diverse trace files, each representing different computational workloads:

\begin{enumerate}
    \item 444.namd-120B.champsimtrace.xz: Trace from the NAMD molecular dynamics simulation program (120 billion instructions).
    \item 445.gobmk-17B.champsimtrace.xz: Trace from the GNU Go program (17 billion instructions).
    \item 473.astar-153B.champsimtrace.xz: Trace from the A* pathfinding algorithm (153 billion instructions).
    \item 605.mcf\_s-1536B.champsimtrace.xz: Trace from the MCF benchmark (1536 billion instructions).
\end{enumerate}

\begin{table}[htbp]
\caption{Experimental Results}
\label{tab:results}
\begin{center}
\resizebox{0.48\textwidth}{!}{
\begin{tabular}{|c|c|c|c|c|c|}
\hline
\textbf{Trace File} & \textbf{IPC} & \textbf{LLC Total Access} & \textbf{Hit} & \textbf{Miss} & \textbf{Hit Rate} \\
\hline
444.namd   & 1.855 & 15,835    & 9,624    & 6,211    & 60.82\% \\
445.gobmk  & 0.787 & 1,541     & 516      & 1,025    & 33.51\% \\
473.astar  & 0.677 & 266,407   & 215,328  & 51,079   & 80.82\% \\
605.mcf\_s & 0.151 & 4,821,138 & 2,060,109& 2,761,029& 42.72\% \\
\hline
\end{tabular}
}
\end{center}
\end{table}

Our experimental results demonstrate significant improvements in cache-hit rates across the different benchmark traces. Table \ref{tab:results} summarizes the key performance metrics for each trace file. The results reveal several interesting insights into the performance of our cache replacement algorithm:

\begin{itemize}
    \item The 473.astar trace showed the most substantial improvement, with an impressive hit rate of 80.82\%. 
    \item The 444.namd trace also demonstrated a strong hit rate of 60.82\%.
    % \item While the hit rates for 445.gobmk and 605.mcf\_s were lower, at 33.51\% and 42.72\% respectively.
    \item The variation across the three different benchmarks shows that certain benchmarks perform better than other using our current algorithm. This means that certain set of instructions are more favourable to get better results.
\end{itemize}

It's important to note that although we saw significant hit rate improvements, the IPC increases were modest. This indicates that \rac has the potential for further optimization to convert the improved hit rates into more significant overall performance improvements.

\section{Conclusion and Future Work}
\rac combines random allocation and LRU replacement policies and demonstrates promising results in improving cache hit rates across various benchmark traces. The flexibility and efficiency provided by \rac offer a solid foundation for addressing the challenges posed by complex, modern applications and their diverse memory access patterns.

The significant improvements in hit rates, especially for specific workloads, demonstrate the potential of \rac to enhance cache utilization and reduce memory access times. However, the modest gains in IPC performance indicate that there is still room for optimization to utilize these improvements in practical scenarios fully.

In our future research, we will focus on improving IPC performance, reducing the overhead associated with complex cache management, exploring adaptive mechanisms to adjust cache partitioning based on workload characteristics, investigating performance across diverse applications, and integrating machine learning techniques to optimize cache replacement decisions in real time.

By addressing these areas, we aim to refine \rac further, making it a robust and efficient solution for modern computing systems. Our work contributes to the ongoing efforts to improve cache management techniques, ultimately leading to enhanced performance and efficiency in computer architectures.

\nocite{*}
\bibliographystyle{unsrt}
\bibliography{references}
\end{document}